\documentclass{elsart3p}
\usepackage{graphicx}
\usepackage{amsmath}

\begin{document}
\begin{frontmatter}

\title{Kaon condensate with
 trapped neutrinos 
and high-density symmetry energy behavior.}
\author[UJ]{Andrzej Odrzywolek}
\author[UJ,IFJ]{Marek Kutschera}

\address[UJ]{M. Smoluchowski Institute of Physics, 
Jagiellonian University, 
Reymonta 4, 30-059 Krakow, Poland}
\address[IFJ]{H. Niewodniczanski Institute of Nuclear Physics, 
Polish Academy of Sciences,
152 Radzikowskiego St, 31-342 Krakow, Poland}

\begin{abstract}
Effects of the neutrino trapping and symmetry energy
behavior are investigated in the framework of the chiral Kaplan-Nelson
model with kaon condensation. Decrease in the condensation threshold
during deleptonization if found to be generic regardless uncertainties in
the nucleon-kaon interactions and symmetry energy. Quantitatively however, 
differences are shown to be important.
\end{abstract}

\end{frontmatter}

\section{Introduction}

Neutron stars are born in core-collapse supernova explosions 
\cite{SN1987Aconf, blondin, janka,  burrows_new, acta, arnett}.
First minutes of their life is a period of rapid evolution \cite{luc}. This is 
protoneutron star (PNS) stage, where matter is transformed to final
cold catalyzed matter \cite{glendenning} state: one of the greatest 
mysteries in astrophysics \cite{Prakash}.
Among other matter with kaon condenstates, proposed by
Kaplan and Nelson \cite{KandN1, KandN2}, is very intriguing possibility.
Unfortunately, existing experimental and observational data
is still unable to select one, correct model \cite{SN1987Aconf}. 
Therefore further investigation of effects present under 
particular assumptionson high-density matter model is in place.

Possibly, only nearby core-collapse supernova explosion
will allow researchers to collect enough data, mainly in neutrino
channel (cf.~Fig.~3 in \cite{lattPRL}), to find clear signatures of particular model.
However, neutrinos are trapped in protoneutron stars. Therefore,
it is required to study nuclear matter with two-parameter 
( baryon number $n_B$ and lepton number $Y_{L}$)
model at least.

Deleptonization in the first seconds after PNS is born
causes decrease in lepton number from value typical
for initial pre-supernova ,,Fe'' matter $Y_{Le}\sim0.4$ \cite{Heger} 
down to some numerically small
value, e.g. $Y_{Le}=0$ in pure neutron matter model. Meanwhile
neutrinos escape outer PNS region, but in central, high-density
core we may assume quasistatic evolution of matter parameterized
with decreasing $Y_{Le}$.

We have dropped all finite temperature effects for simplicity,
however they are potentially equally important \cite{Prakash}.
Generally, effects of decreasing temperature are smaller than
decreasing lepton number and act in similar direction on critical 
kaon condensation density.

\section{Model with kaon condensate}

Energy density for matter with kaon condensate is given by: 
\begin{eqnarray} \label{energy} \varepsilon
(n_n, \,n_p, \,\theta , \,{\mu_{e}}, \,{\mu
  _{{\nu _{e}}}}, \,{\mu_K}, \,\mu_\mu, \,{\mu_{{\nu _{
 \mu }}}})= \nonumber\\  
{\varepsilon_{{F_{n}}}} + {\varepsilon _{{F_{p}}}} + {
 \varepsilon
_{{F_{e}}}} + {\varepsilon _{{F_{{\nu _{e}}}}}}
  + {\varepsilon _{{F_{\mu
}}}} + {\varepsilon _{{F_{{\nu _{\mu }}
 }}}}+ {  \varepsilon _{\mathit{int}}}
+{\varepsilon _{\mathit{kaon}}}  
\end{eqnarray}
 
where variables on the left hand are: 
$n_n$, $n_p$ -- neutron and proton density, 
 $\theta$-- amplitude of the kaon condensate 
 $\mu_i$ -- chemical potentials. Right hand
 has been decomposed into:
 ${\varepsilon _{{F_{i}}}}$-- Fermi sea energies,  
${\varepsilon _{\mathit{Kaon}}}$ -- 
kaon-nucleon interaction and
${\varepsilon _{\mathit{int}}}$
  --  nucleon interaction including symmetry energy. 

Nucleon part can be rewritten as:

\begin{equation}
\varepsilon (u, \,x)
=
\frac {3}{5}
E_F^0\,{n_0}\,u^{5/3}
+
u\,{n_0}\,(1 - 2\,x)^{2}\,\mathrm{S}(u)
\end{equation}

where $E_F^0=36.885$ MeV. Standard parameterization 
($n_0=0.16 fm^{-3}$) has been used:
\begin{eqnarray} 
\label{parametrization}
n_p=x\,{n_{B}}
 \\
n_n=(1 - x)\,n_B  
\label{parametrization1}
 \\
n_B=u\,n_0 
\label{parametrization2}
\end{eqnarray}
and $S(u)$ is the {\em symmetry energy}.

Leptons contribute to the total energy:
\begin{equation}
\label{lepton_sea}
\varepsilon_{F_i}=
\frac{g_i \;{\mu_i}^4}{8\,\pi^2},
\end{equation}
where $g_e$=2 and $g_{\nu}=1$ (see, however,
footnote \ref{fn_osc}).

Generalization to other lepton families is trivial. However,
we have not included $\mu$ and $\tau$
flavors, as their contribution is minimal \cite{Thorson,KK}.

Contribution to the total energy from the kaon condensate
can be computed from  the  Kaplan--Nelson chiral formalism  \cite{Fuji, Thorson}:
\begin{eqnarray} 
\label{wzornakaony}
\varepsilon_{kaon} =  \frac{\mu^2\,f^2\,\sin^2{\theta}}{2}  
+(\cos{\theta} -1) \times \\ \nonumber
\times \left[
n_B \, x \, \Sigma_{Kp} 
+ n_B \,(1 - x)\,\Sigma_{Kn} 
- 
f^2 \,m_K^2
\right]  
\end{eqnarray}
where,
$m_K=493,7$~MeV and  
$f=93$~MeV is  kaon mass and pion decay rate, respectively. Constants  
$\Sigma_{Kp}$ and  
$\Sigma_{Kn}$ 
define kaon-proton and kaon-neutron interaction strength.
Due to large uncertainties they may be replaced with single
quantity $\Sigma _{\mathit{KN}}$ placed between   
  $168$ and $520$MeV.
Some authors use $a_{3}m_{s}$ instead:
\begin{equation} 
\label{sigmatoa3ms}
\Sigma_{KN}=-\frac{1}{2}\,(a_{1}m_{s}+2\,a_{2}m_{s}+4\,a_{3}m_{s})
\end{equation}
where $a_{1}m_{s}=-67$~MeV, $a_{2}m_{s}=134$~MeV.

Kaon density is \cite{Prakash}:
\begin{equation} \label{n_kaon}
n_K = {\mu _{K}}\,f^2 \, \sin^2{\theta} + {n_{B}}\, 
\left( \frac {1}{2}\,x + \frac {1}{2} \right) \,( 1 - \cos{\theta} )
\end{equation}

Minimalization of the energy must include conservation
of the baryon number, electric charge and lepton numbers
if the neutrinos are trapped:
\begin{subequations}
\begin{equation} 
{n_{n}} + {n_{p}}={n_{B}} 
\end{equation}
\begin{equation} 
{n_{p}}={n_{e}} + {n_{K^{-}}} + {n_{\mu }} 
\end{equation}
\begin{equation}
\label{leptonnumbers}
x_{i} + Y_{\nu_i}=Y_{Li}
\end{equation}
\end{subequations}
Baryon number density is a free parameter. Charged lepton 
fractions are denoted as $x_i\equiv n_i/n_B$, neutrinos as $Y_{\nu_i}$. 
Conservation of the electric charge is ensured using kaon chemical potential as
a Lagrange multiplier:
\begin{equation} 
\label{Q}
\tilde{\varepsilon}=\varepsilon  - \mu \,({n_{p}} - {n_{e}
} - {n_{K^{-}}} - {n_{\mu }}).
\end{equation}

Electroweak and strong interactions allow reactions:
\begin{subequations}
\begin{equation}
n\longleftrightarrow p^{+} + e^{-} + \bar{\nu} _{e}
\end{equation}
\begin{equation} 
n\longleftrightarrow p^{+} + K^{-}
\end{equation}
\begin{equation}
n\longleftrightarrow p^{+} + \mu^{-}  + \bar{\nu} _{\mu }
\end{equation}
\end{subequations}
and respective chemical potentials obey:
\begin{subequations}
\label{eq_nu}
\begin{equation}
{\mu _{n}}-{\mu _{p}} = {\mu _{e}} - {\mu _{{\nu _{e}}}} 
\label{eq_nu_1}
\end{equation}
\begin{equation}
{\mu _{n}}-{\mu _{p}} = {\mu _{\mu }} - {\mu _{{\nu _{\mu }}}} 
\label{eq_nu_2}
\end{equation}
\begin{equation}
{\mu _{n}}-{\mu _{p}} = {\mu _{K}} 
\label{eq_nu_3}
\end{equation}
\end{subequations}

If neutrinos escape freely, one can put simply \mbox{$\mu_{\nu_i}\equiv0$}
and we have left with only one independent chemical potential equal
for all negative electric charge particles.

\section{Kaon condensation without neutrinos \label{RozdzSM} }

If neutrinos escape freely (old neutron star case)
we have only one driving parameter: baryon density.
Energy is minimalized numerically solving system of 
equations:
\begin{equation}
\label{urnumeric}
\frac{\partial \,\tilde{\varepsilon} (x, \,\mu  
, \,\theta )}{\partial \theta }=
\,{\frac {\partial\,\tilde{\varepsilon} (x 
, \,\mu , \,\theta ) }{\partial x}}=
 \,{\frac {\partial\, 
\tilde{\varepsilon} (x, \,\mu , \,\theta ) }{\partial \mu }}=0
\end{equation}

Functions $x(u), \,\theta(u), \,\mu(u)$ are immediate result of calculations. 
Other properties, e.g.  ${n_{e}}(u), \,{n_{K}}(u)$ and EOS
 can be then easily obtained.
Condensation threshold is defined as a maximum density
where still $\theta (u)=0$. 
Numerical results are presented for $u$ up to 12 for
three values ${\Sigma
_{\mathit{KN}}}=168,\,344,\,520$~MeV\footnote{Equivalent values are: 
$a_3\,m_s=- 134,\,- 222,\,-310$~MeV}: values covering entire considered 
range of  values for this parameter.

Larger $\Sigma_{KN}$ (i.e smaller $a_{3}m_{s}$) gives
stronger kaon-nucleon interaction and lower condensation threshold
(Fig.~\ref{figure_3}). Both amplitude of the condensate and proton fraction
(Fig.~\ref{figure_2}, \ref{figure_3}) exhibit asymptotic behavior.
This is typical if symmetry energy is constant or growing with density.
 Chemical potential for kaons and electrons (Fig.~\ref{figure_4}) 
can reach large ($\mu<100$~MeV) negative values  if $\Sigma_{KN}$ 
is large, and muon flavor will be produced. Overall contribution from  muons
is however small \cite{KK, Thorson}.
As lepton number is not conserved electron/positron fraction
can be relatively high. This behavior is
completely changed with neutrino trapping, as we explain in the next section.

\section{Kaon condensate and neutrino trapping}

If neutrinos are trapped then properties of the dense matter
are numbered by the two parameters: baryon density $n_B$
and lepton number density $Y_{Le}$. In principle we have three
separate lepton numbers, but initially only $Y_{Le}$
is not identically zero. Therefore, due to large muon and taon
masses we may safely restrict to electron lepton number conservation 
alone\footnote{
\label{fn_osc}
Neutrino oscillation phenomenon indicates
conservation of the total lepton number only.
Therefore \eqref{leptonnumbers} could be replaced
with $x_e+ Y_{\nu_e}+Y_{\nu_\mu}+Y_{\nu_\tau} = Y_L \equiv Y_{Le}$,
where we have put $Y_L\mu=Y_L\tau=0$.
As neutrinos have very small masses all three terms are identical
and finally we get  $x_e+ 3 Y_{\nu_e} = Y_{Le}$. Only difference
is ''degeneracy factor'': $g=3$ instead of $g=1$ in eq.~\eqref{lepton_sea}
for neutrinos.
}.

Lepton number conservation can be introduced into 
eq.~\eqref{Q} in the following manner. We rewrite
\eqref{leptonnumbers} using chemical potentials:
\begin{equation} 
\label{Uklad2}
{\mu _{e}}^{3} + \frac {1}{2}\,{\mu _{{\nu
_{e}}}}^{3}=3\,\pi   ^{2}\,{n_{B}}\,{Y_{\mathit{Le}}}.
\end{equation}
Eq.~\eqref{eq_nu_1} minus  \eqref{eq_nu_3} gives: 
\begin{equation} 
\label{Uklad1}
{\mu_{K}}={\mu_{e}}-{\mu_{{\nu}_{e}}}
\end{equation}

Eq.~\eqref{Uklad1} and \eqref{Uklad2} are used to derive
$\mu_e$ and $\mu_{\nu_e}$ as a functions of $\mu_K$, $n_B$
and $Y_{Le}$. Now, minimalized function is:   
\begin{eqnarray*} 
\label{efenergynu} \nonumber
\tilde{\varepsilon} ({Y_{\mathit{Le}}}, \,{n_{B}}, \,x, \,\theta , \,{
\mu _{K}})={\varepsilon _{{F_{n}}}} + {\varepsilon _{{F_{p}}}} 
+ 
\varepsilon _{\mathit{Kaon}}
\\  
+ {\varepsilon _{{F_{{\nu _{e}}}}}}
({\mu _{K}}, \,{Y_{\mathit{Le}}}) 
- \mu_K\,
\left[ n_{p} - n_{K
} - {n_{e}}({\mu _{K}}, \,{Y_{Le}})
\right ]
\end{eqnarray*}
Expression above is {\em explicite} very complex due to presence
of the third order radicals resulting from \eqref{Uklad2}.

\begin{figure}
\includegraphics[width=0.9\columnwidth]{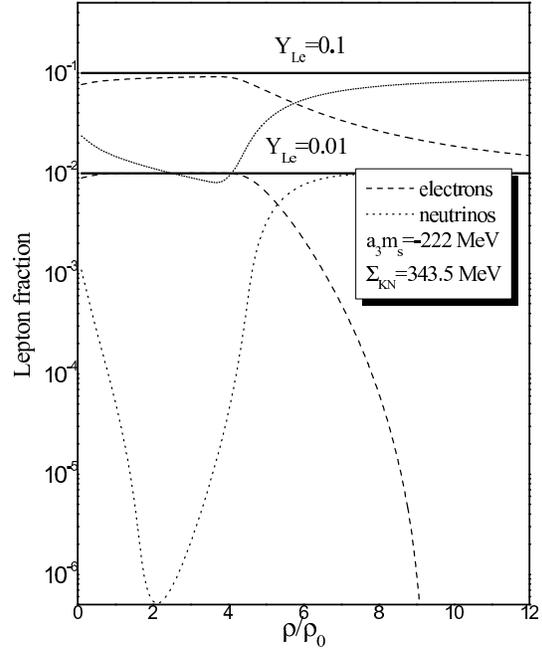}
\caption{
\label{figure_1}
Leptons fraction {\em versus} baryon density. 
}    
\end{figure}

\begin{figure} 
\includegraphics[width=0.9\columnwidth]{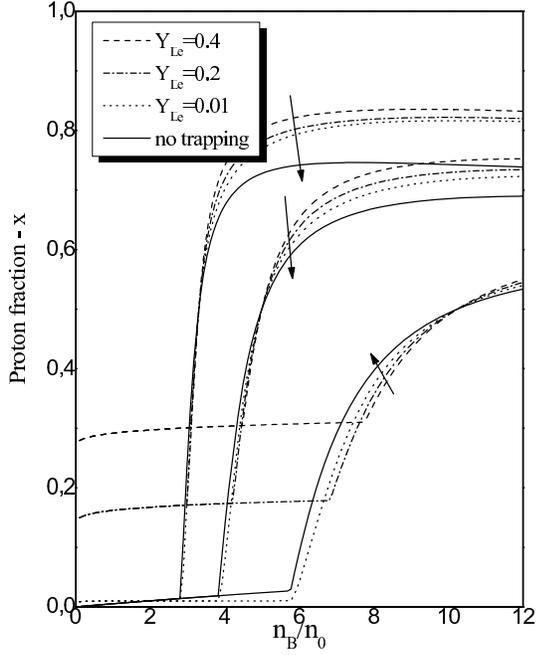}
\caption{
\label{figure_2}
Proton fraction for three values of the $\Sigma_{KN}$. Arrows
indicate direction of the deleptonization effects.
}  
\end{figure}

Two parameter family of solutions is presented 
in Fig.~\ref{figure_5}.
Proton fraction strongly depends on trapped lepton number 
until kaon condensation threshold. This behavior
depends somewhat on $\Sigma_{KN}$, cf.~Fig.~\ref{figure_2}.
Fig.~\ref{figure_1} illustrate lepton number conservation. Without
kaons electrons are required by electric charge conservation.
If condensate is present, negative charge is provided
by preferred due to strong interactions kaons, 
and neutrinos begins to provide required lepton number amount.

\begin{figure}
\centering
\includegraphics[width=0.9\columnwidth]{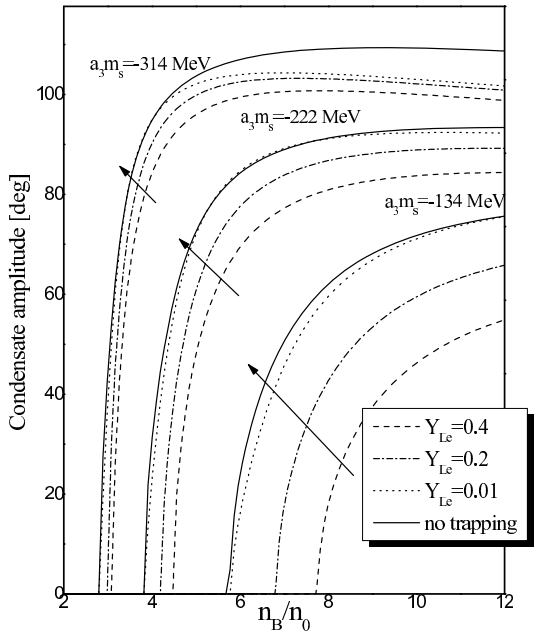}
\caption{
\label{figure_3}
Deleptonization effects on the amplitude of the
kaon condensate $\theta$. Threshold for condensation
with trapped neutrinos is always higher and proportional
to the lepton number density $Y_{Le}$.   
} 
\end{figure}

The most important effect of deleptonization is decrease in
kaon condensation threshold, cf.~Fig.~\ref{figure_3}. Direction
of the effect do not depend on $\Sigma_{KN}$. Therefore,
as kaons tends to ,,soften'' EOS, deleptonization
cause decrease in maximum neutron star mass. If PNS is born
in stable state with large $Y_{Le}$ then deleptonization
may cause delayed collapse to a black hole if $Y_{Le}$
reach small values long time (i.e. tens of seconds \cite{acta}) 
after core-collapse\footnote{Neutron star has not been found
in the remnant of the supernova 1987A \cite{Suntzev_talk}. 
Delayed collapse is the most probable explanation.}.    
 
\begin{figure} 
\includegraphics[width=0.9\columnwidth]{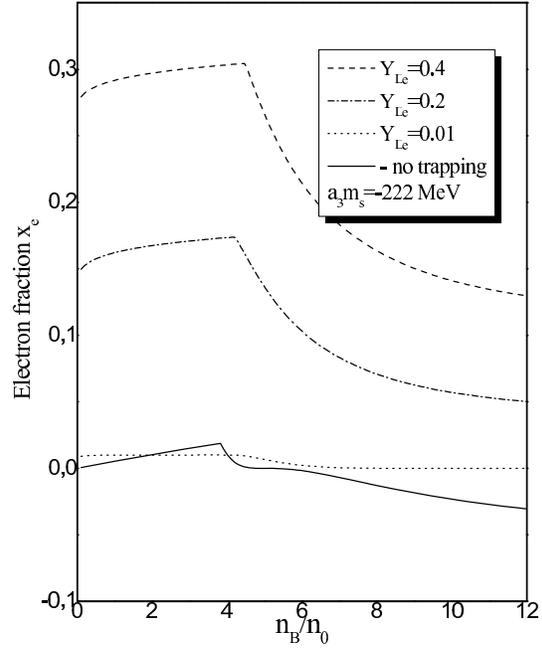}
\caption{
\label{figure_4}
Electron fraction with and without neutrino trapping.
}  
\end{figure}

\begin{figure} 
\centering
\includegraphics[width=0.9\columnwidth]{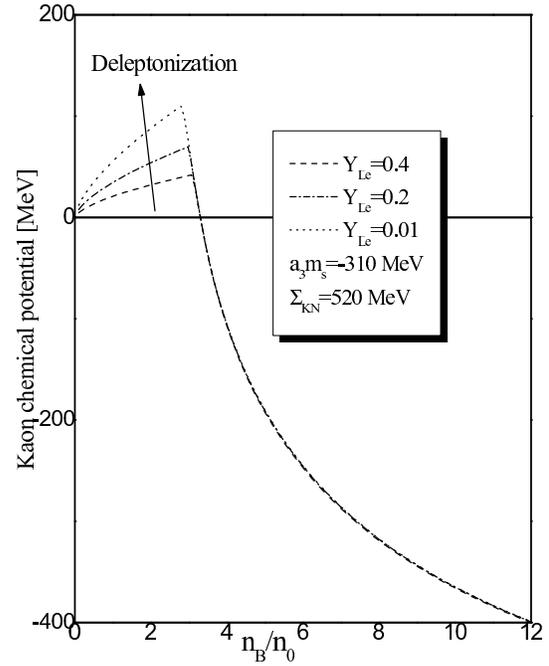}
\caption{
\label{figure_5}
Kaon chemical potential as a function of nuclear density.
}
\end{figure}

Model with $Y_{Le} \to 0$ is clearly different
from model without neutrino trapping (Fig.~\ref{figure_4}). Initially,
for large values of $Y_{Le}$ model works well, but transition to the
free streaming regime requires solving transport equations rather
than quasistatic evolution.

\section{Symmetry energy effects \label{RozdzialSym}} 
 
High density energy symmetry behavior is very important for neutron stars
and core-collapse supernovae (Lattimer \& Yamada at \cite{SN1987Aconf}) 
but is a source of the great uncertainty
\cite{KK, KK1, KK2, KK3}. Two qualitative behavior has been found: 
\begin{itemize}
\item{always growing (from mean field theories) }
\item{decreasing at high densities (variational methods)}
\end{itemize}

Mean field theory results can be approximated with \cite{Thorson}:
\begin{equation}
V_2 = a \, u, \; V_2 = a \, \sqrt{u},\; V_2 = a \, \frac{2\,u^2}{1+u} 
\end{equation}
where $a=17$~MeV.

\begin{figure}
\includegraphics[width=0.9\columnwidth]{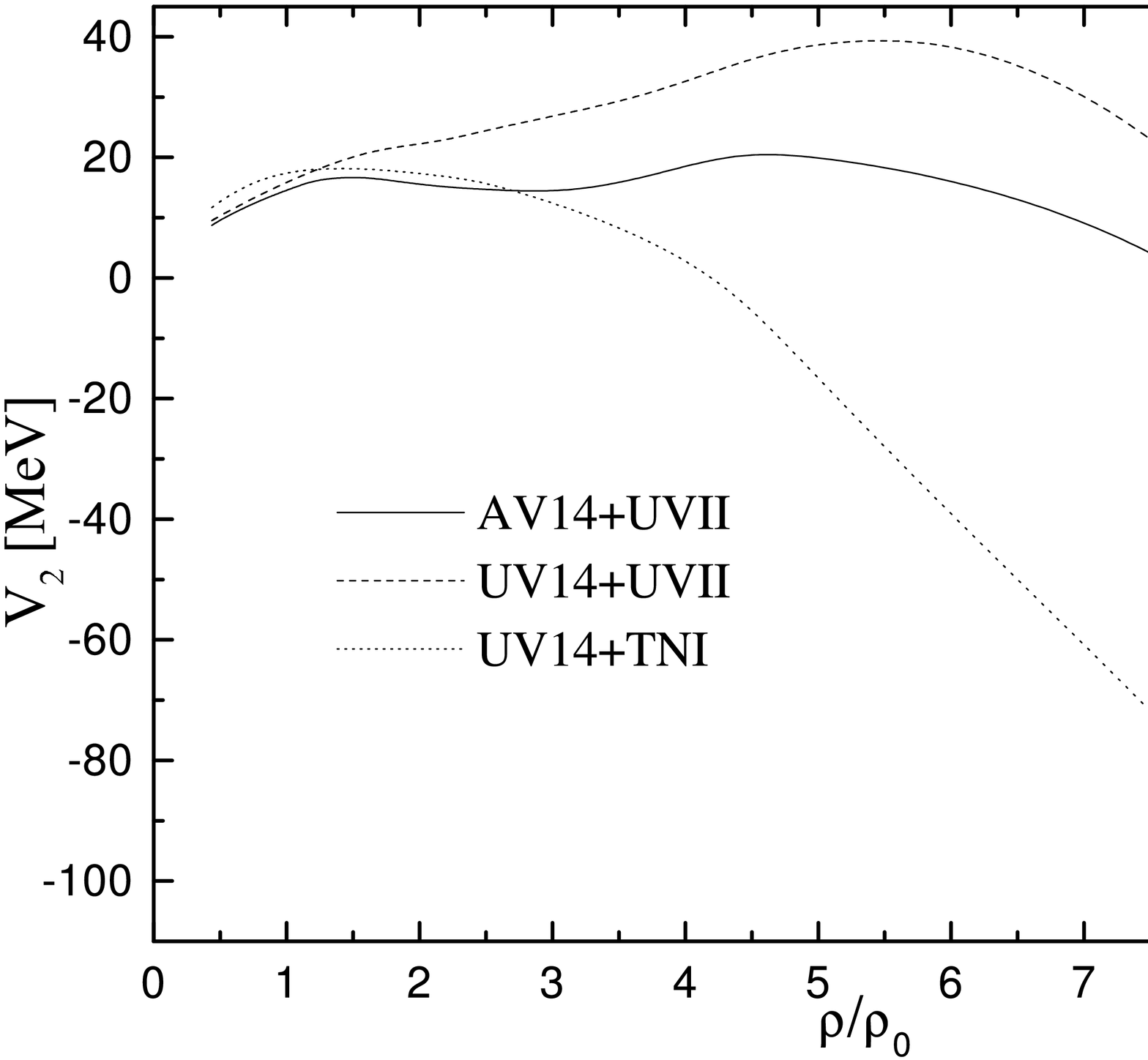} \\
\includegraphics[width=0.9\columnwidth]{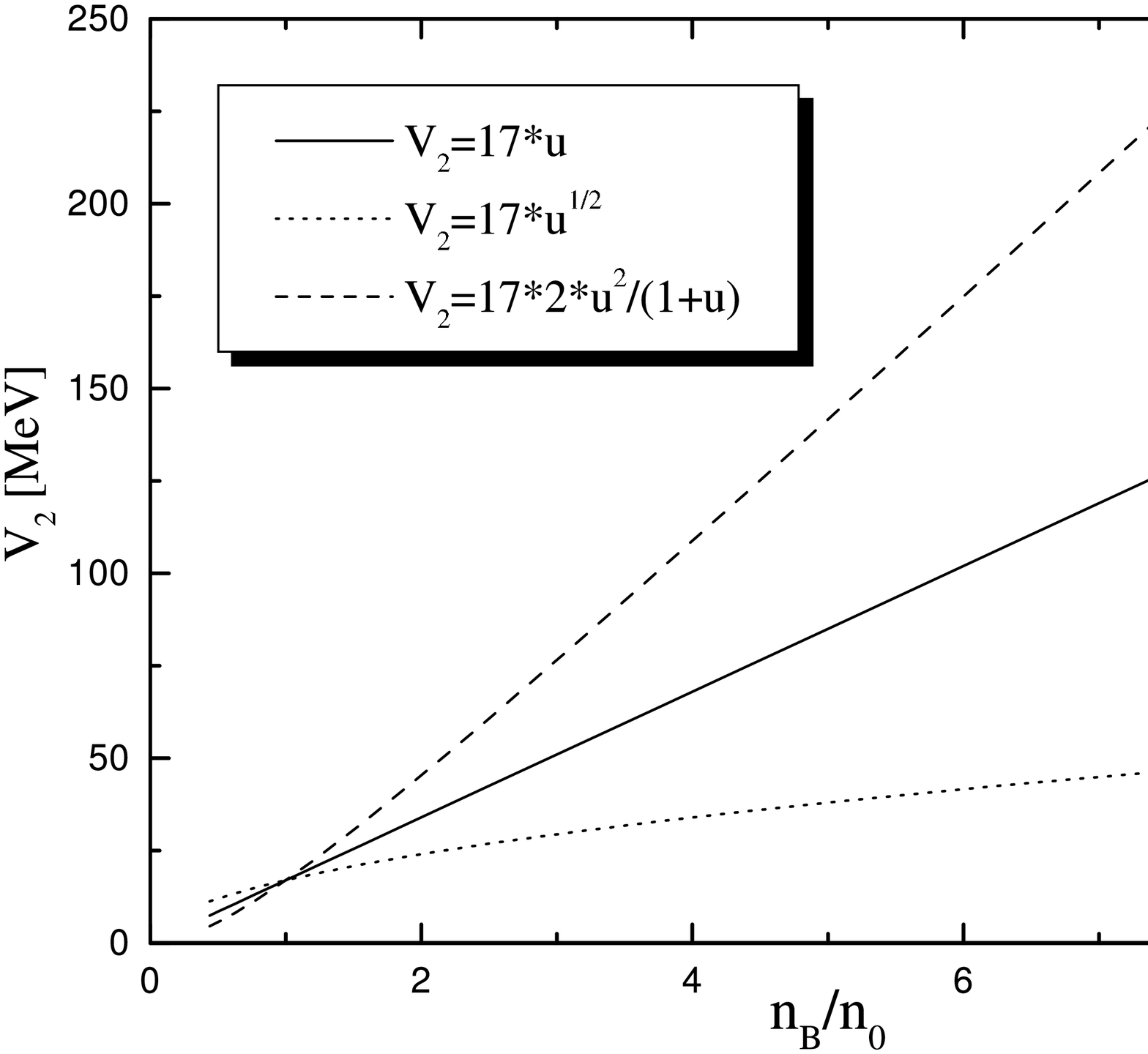}
\caption{
\label{figure_6}
Symmetry energy in variational models (upper) and mean field models (lower). 
Results are presented for UV14+UVII and linear cases (solid lines). 
}
\end{figure}

Fig.~\ref{figure_6} compare mean field and variational results.
 In the latter, symmetry energy not only decrease, but can reach negative
 values and pure proton or pure neutron states are preferred.

\begin{figure}
\includegraphics[width=0.9\columnwidth]{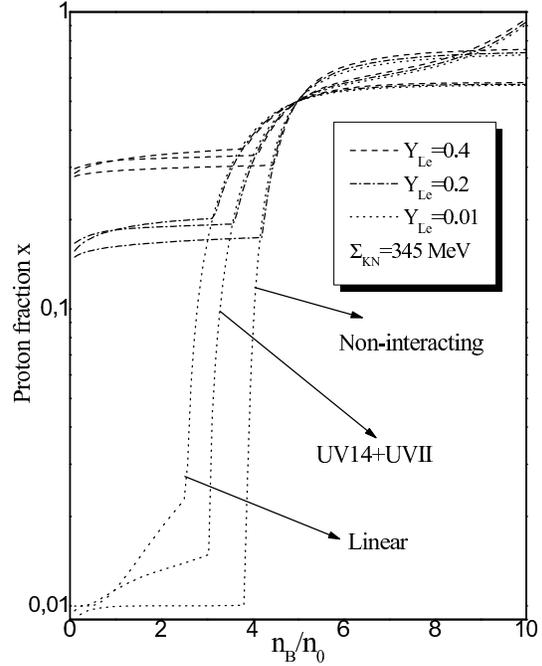} 
\caption{
\label{figure_7}
Deleptonization effects on proton fraction for three
symmetry energy models for  $a_{3}m_{s}=-222$~MeV. It is clear
that uncertainty due to symmetry energy leads
to larger effects than deleptonization itself. 
}  
\end{figure}

\begin{figure} 
\centering
\includegraphics[width=0.9\columnwidth]{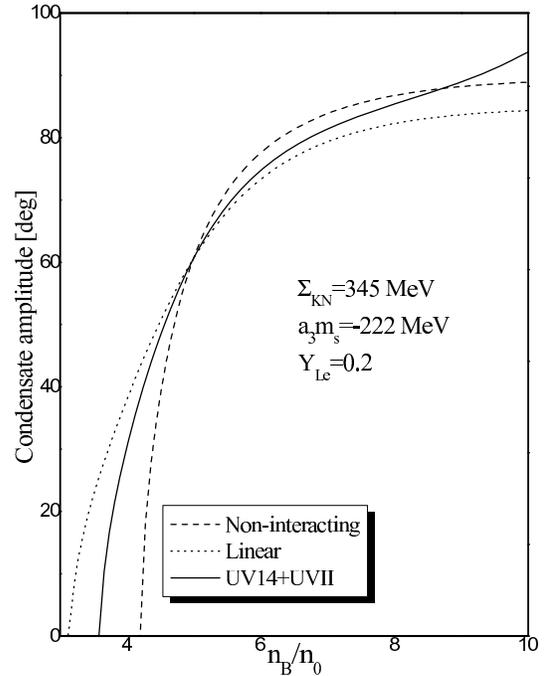}
\caption{
\label{figure_8}
Kaon condensate amplitude for $Y_{Le}=0.2$ 
in various symmetry energy models. 
}
\end{figure}

\begin{figure}
\centering
\includegraphics[width=0.9\columnwidth]{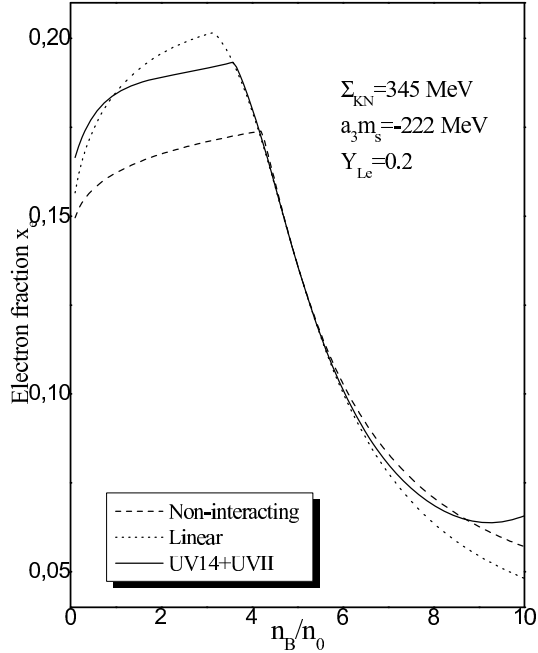}
\caption{The same as in Fig.~\ref{figure_8} for the electron fraction.} 
\label{figure_9}
\end{figure}

Symmetry energy strongly influences on matter properties both
below and above condensation threshold (Figs.~\ref{figure_7}-\ref{figure_9}).

\begin{figure}
\includegraphics[width=0.9\columnwidth]{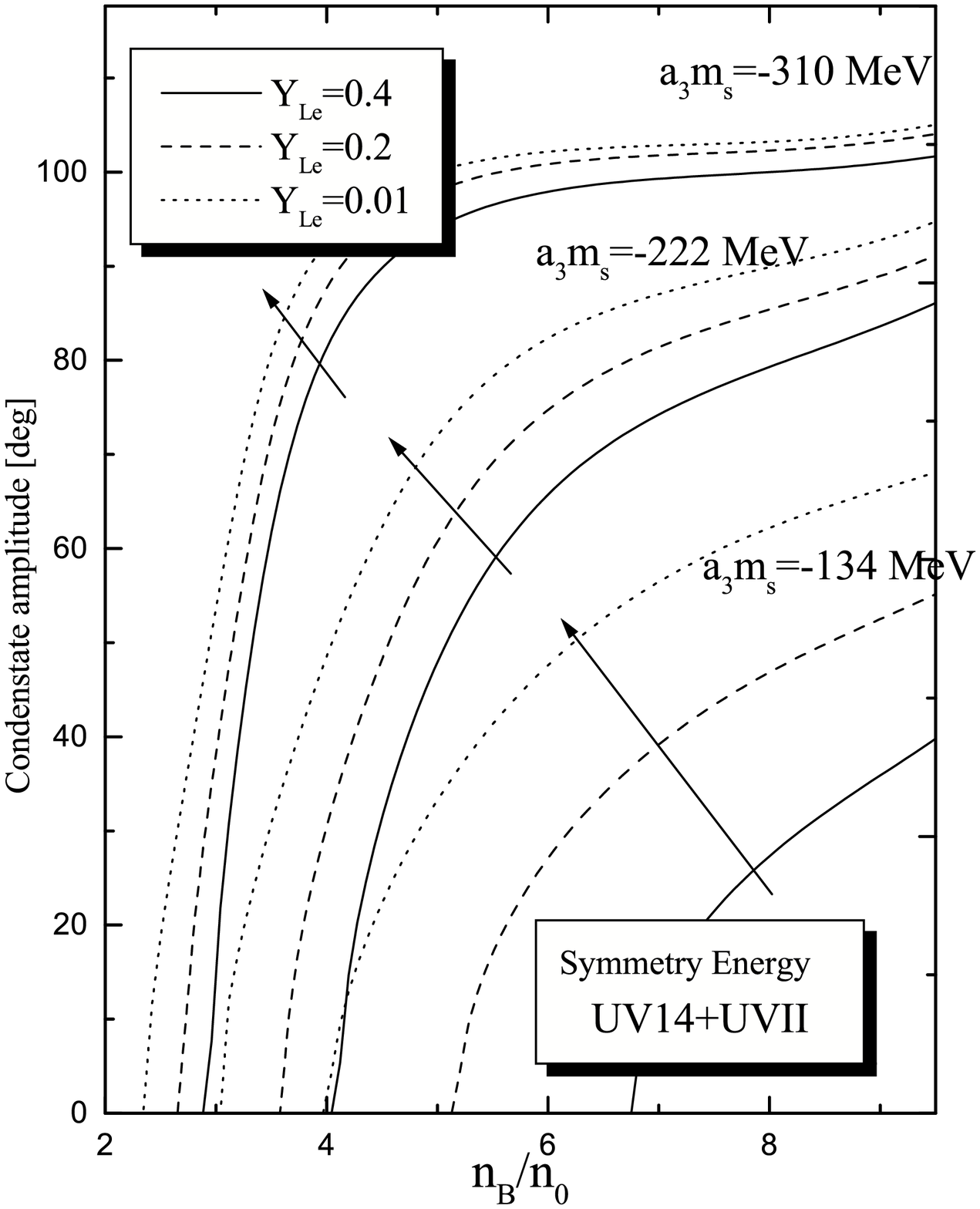}
\caption{
\label{figure_10}
Deleptonization effect on kaon condensate with symmetry energy are qualitatively
similar to the case without symmetry energy, cf. Fig.~\ref{figure_3}.
}  
\end{figure}

General tendency to decrease condensation threshold
is however unaffected, and amplitude is still growing 
with density (Fig.~\ref{figure_10}). 

\section{Conclusions}

Decrease of the kaon condensation threshold during deleptonization 
has been shown to be universal in the considered class of models. 
Main source of the uncertainty is the high density
behavior of the symmetry energy and value of the kaon-nucleon interaction
parameter $\Sigma_{KN}$. Decrease in the condensation threshold
may cause newborn PNS to be unstable as trapped lepton number
is carried away from the neutrinospheres \cite{Prakash}.

Increasing condensate volume finally may cause collapse
to a black hole and immediate disappearance of the neutrino flux 
(effect observable for a next Galactic supernova) as neutrino spheres
are swallowed under event horizon \cite{Beacom}.

This work was supported by grant of the Polish  Ministry of 
Education and Science No. 1~P03D~005~28.

\end{document}